\documentclass[onecolumn]{emulateapj}
\usepackage{natbib}
\usepackage{psfig}
\begin{document}

\renewcommand{\thefootnote}{\fnsymbol{footnote}}

\newcommand{\qso}{QSO 2237+0305}

\title[{\it Spitzer} Observations of \qso]{{\it Spitzer} 
observations of a gravitationally lensed quasar, \qso}
\author{Eric Agol$^1$, Stephanie M Gogarten$^1$, 
Varoujan Gorjian$^2$, Amy Kimball$^1$}

\address{$^1$ Department of Astronomy, University of Washington,
  Box 351580, Seattle, WA 98195, 
agol@astro.washington.edu}
\address{$^2$ Jet Propulsion Laboratory, California Institute of
  Technology, 4800 Oak Grove Drive, Pasadena, CA 91109}

\begin{abstract}
The four-image gravitationally lensed quasar \qso\ is microlensed
by stars in the lens galaxy.  The amplitude of microlensing variability
can be used to infer the relative size of the quasar as a function
of wavelength; this provides a test of quasar models.  Towards this end, we 
present \emph{Spitzer Space Telescope} Infrared Spectrograph (IRS) 
and Infrared Array Camera (IRAC) observations of \qso, finding the 
following: (1) The infrared spectral energy distribution is similar 
to that of other bright radio-quiet quasars, contrary to an earlier 
claim.  (2) A dusty torus model with a small opening angle fits the 
overall shape of the infrared spectral energy distribution well, but 
the quantitative agreement is poor due to an offset in wavelength of 
the silicate feature.  (3) The flux ratios of the four lensed images 
can be derived from the IRAC data despite being unresolved.  We find 
that the near-infrared fluxes are increasingly affected by 
microlensing towards shorter wavelengths. (4) The wavelength dependence 
of the IRAC flux ratios is consistent with the standard quasar model 
in which an accretion disk and a dusty torus both contribute near 1 
micron in the rest frame.  This is also consistent with recent infrared 
spectropolarimetry of nearby quasars.
\end{abstract}

\section{Introduction}

Radio quiet quasars, or  quasi-stellar objects (QSOs), are some of the most 
luminous objects in the Universe, reaching 10$^{13-14} L_\odot$ in the brightest
cases; they are also very compact, hence the name ``quasi-stellar."  Such a 
large luminosity from a compact source cannot be powered by stars, but
can be powered by a super-massive black hole at the center
of a galaxy \citep{LyndenBell1969}.  The black hole creates radiation by 
accreting gas via an accretion disk near the Eddington limit.  The accretion 
disk is fed by gas from the surrounding galaxy via a dust and gas torus on 
parsec scales.  This widely held picture explains 
the two most significant features of quasar spectral energy distributions: 1) a 
broad peak in the optical/ultraviolet due to the accretion disk and 2) a broad 
peak in the infrared due to the dusty torus.  These two spectral components are 
commonly referred to as the ``Big Blue Bump" \citep{Shields1978} and 
the ``infrared bump" \citep{Sanders1989}, with comparable luminosities in each.   
In between these two peaks lies a valley dubbed ``the one-micron dip."
The standard model naturally accounts for the one-micron dip due to the sublimation 
temperature of dust; the dusty torus is heated by radiation from the accretion
disk, but dust cannot exist at temperatures above about 1500 $K$, causing a
cutoff in the emission from the torus that always occurs near 1 micron.
This paper presents a novel test of this two-component model using measurements
of gravitational microlensing near the one micron dip of the high redshift quasar 
\qso\ ($z_s = 1.695$).  Near one micron both the accretion disk and dusty torus 
have nearly equal specific luminosity, but very different sizes, so
the region near one micron is ideal for testing the standard model with microlensing.

\subsection{Background}

\qso\ was chosen for this study as it holds several records among 
gravitationally lensed quasars: it was one of the first four-image lenses 
discovered \citep{Huchra1985}; its lens galaxy has the lowest redshift, $z_l=0.0395$ 
\citep{Huchra1985}; and it was the first to show 
gravitational microlensing \citep{Irwin1989}.  This last fact is a result
of the second:  a nearby lens galaxy causes a large
velocity of the quasar relative to the magnification patterns created by stars
in the lens galaxy projected onto the source plane; this large relative
velocity results in a shorter timescale for microlensing.  The discovery of 
microlensing in this system and its short microlensing time scale made it a 
``rosetta stone" for studies of the size of the quasar emission
region:  the time-dependent microlensing magnification is sensitive to the 
size of the source, effectively resolving the quasar on sub-microarcsecond 
scales.  Larger sources smooth over the microlensing magnification pattern and 
thus experience smaller and more gradual variations in magnification 
\citep{Refsdal1991}.  
In unlensed quasars, only the spectral energy distribution 
can be compared to models \citep[e.g.][]{Sanders1989,Blaes2001,Malkan1983},
while for \qso\ the size as a function of wavelength can be compared to models 
as well, in principle giving much stronger constraints on the emission mechanism.

Despite this promise, the interpretation of the first optical microlensing 
events in \qso\ were puzzling: one study showed the inferred size of the Big 
Blue Bump was consistent with the accretion disk model \citep{Wambsganss1990}, 
while another study showed the size of the emission region was too {\it small}
\citep{Rauch1991}.  The latter result led to other newer models which require
more theoretical development, e.g.\ \citet{Barvainis1993,Czerny1994}.
With a much larger data set and more sophisticated analysis of the
microlensing lightcurves, \citet{Kochanek2004} showed that thermal emission
from an accretion disk is consistent with the size inferred from
microlensing.  However, microlensing in a sample of gravitationally lensed
quasars has led to a different conclusion: the size of the
optical/UV emission region inferred from microlensing is too {\it large}
compared with the size of quasar accretion disk models inferred from
fitting the SEDs \citep{Pooley2007,Morgan2007}.  

This confused state of affairs of microlensing of the Big Blue Bump partly 
stems from the fact that the {\it absolute} size is difficult to constrain 
as there are degeneracies between the mass of the microlenses and the sky 
velocity of the quasar relative to the magnification pattern.  However, the 
{\it relative} size versus wavelength is much easier to constrain since it is 
not as subject to these degeneracies \citep{Wambsganss1991}.  In particular, 
the first results for the
wavelength-dependent relative size seem to be in good agreement with the 
accretion disk model for a different lensed quasar \citep{Poindexter2008}, 
although the absolute size is still discrepant.  The relative size 
of the optical/ultraviolet/X-ray emission region for \qso\ is well 
constrained by microlensing \citep[e.g.][]{
Wyithe2000,Kochanek2004,Anguita2008}, taking advantage of the long
time scale data set collected by the Optical Gravitational
Lensing Experiment \citep[OGLE][]{Udalski2006,Wozniak2000}.
An intensive monitoring campaign with the Very Large Telescope (VLT) promises to 
give a very detailed picture of the relative sizes of the Big Blue Bump 
and broad-line regions as a function of wavelength \citep{Eigenbrod2008a,
Eigenbrod2008b}.  In this paper we will not attempt to resolve the
absolute size problem, but rather we will argue that the standard 
two-component model provides good agreement with the wavelength 
dependence of the microlensing flux ratios, adding credence to 
the standard model.

The first microlensing study of the infrared bump was carried out with 
\qso\ to distinguish synchrotron and dust emission 
models for the infrared bump \citep{Agol2000}.  The synchrotron emission
model provides an alternative, albeit less natural, explanation for the
infrared bump.  The synchrotron emission region responsible for the 
infrared bump has to be compact to avoid self-absorption; thus it should 
show strong variability due to microlensing.  On the other hand the dusty 
torus model must be extended 
to avoid sublimation, and thus should vary weakly due to microlensing.
\citet{Agol2000} found that the mid-infrared flux ratios were consistent
with no microlensing \citep{Schmidt1998}; this despite
the fact that the optical source was simultaneously undergoing strong 
microlensing events.  These observations ruled out strong microlensing 
magnification of the mid-infrared emission region, which was one of the 
first clear-cut demonstrations that the infrared emission region in 
radio-quiet quasars is due to thermal emission by dust, not synchrotron 
emission \citep{Wyithe2002}.   Here we extend these results to
observations near the 1 micron dip where both the dusty torus and
accretion disk contribute to the flux.

\subsection{Plan of the paper}

In \S 2 we discuss the observations and data reductions.
Although the primary focus of this paper is on probing the relative
source size versus wavelength, there are two problems related to
the SED that may be addressed with our data as well: 1) How similar 
is the SED of \qso\
to other quasars and Seyfert galaxies?  Ground-based observations
indicated that \qso\ contained hotter dust than any other quasar
\citep{Agol2000}, while the observations presented here show that 
the ground-based observations at one wavelength were in error.  The 
large intrinsic 
luminosity and high magnification, $\mu \sim 16$, \citep{Schmidt1998} make 
this QSO an excellent candidate for spectroscopy and allow us to compare 
the spectrum of a high-redshift quasar with nearby Seyfert galaxies.
In \S 3.1 we show that \qso\ looks very similar to other quasars and 
Seyfert galaxies.  2) How well does the infrared SED match dusty torus 
models?  In \S 3.2 we show that the overall shape agrees qualitatively,
but the quantitative agreement is poor.

In \S 3.3-3.4 we present the microlensing results and interpretation
for \qso, demonstrating that two size scales are required to
fit the microlensing flux ratios, as predicted by the accretion disk/dusty
torus model. In \S 4 we discuss the implications for quasars in
general and in \S 5 we summarize.

\section{Observations}

Cycle 2 observations with the {\it Spitzer} Space Telescope were awarded
for studying \qso\ under program 20475.
\qso\  ($\alpha = 22^{h}40^{m}30.2^{s}, \delta = 3^{\circ}21'31\farcs1,$
J2000) was observed with both the Infrared Spectrograph
\citep[IRS;][]{Houck2004} and the Infrared Array Camera
\citep[IRAC;][]{Fazio2004} on \emph{Spitzer}.  A summary of observations
is presented in Table~\ref{tbl:obs_sum}.  Listed integration times are
for observations of the QSO only; peak-up observations and sky
observations are not included.

\begin{table}
\caption{\label{tbl:obs_sum}Summary of Observations.}
\centerline{
\begin{tabular}{lcccc}
\hline
Date & Instrument & Module & Integration Time / Exp. & Num. of Exp. \\
\hline
2005-11-17 & IRAC & 3.6 $\mu$m & 1.2 sec & 32\\
2005-11-17 & IRAC & 4.5 $\mu$m & 1.2 sec & 32\\
2005-11-17 & IRAC & 5.8 $\mu$m & 1.2 sec & 32\\
2005-11-17 & IRAC & 8.0 $\mu$m & 1.2 sec & 32\\
2005-11-20 & IRS & Short-Low & 6.29 sec & 64\\
2005-11-20 & IRS & Long-Low & 14.68 sec & 128\\
2006-06-29 & IRS & Long-High & 60.95 sec & 60\\
\hline
\end{tabular}}
\end{table}


\subsection{IRS}

Basic Calibrated Data (BCD) images were obtained from the
\emph{Spitzer} archive, pipeline version S14.0.0.  \qso\ was observed
with the IRS modules Short-Low (SL2, 5.2-8.7 $\mu$m, and SL1, 7.4-14.5 $\mu$m),
Long-Low (LL2, 14.0-21.3 $\mu$m), and Long-High (LH, 18.7-37.2
$\mu$m), for a full observed
wavelength coverage of 5.2-37.2 $\mu$m. 
Rogue pixels were eliminated using the IRSCLEAN\_MASK
software package provided by the \emph{Spitzer} Science Center. 
We created our own rogue pixel maps (pixels with anomalous behavior) for each
spectral order by measuring 
two quantities from a series of sky images for each order (for the long-high
data we chose only portions of the image that did not contain the target): 
(1) the scatter in each pixel with time; (2) the difference between the 
value of a pixel and the median of a region within a 5$\times$5 pixel 
region surrounding it.  We then flagged pixels which had either excessive 
scatter or consistently had values much larger than the median smoothed image, 
and included these in the rogue pixel map.  This procedure resulted in similar
maps to those generated automatically by the IRSCLEAN\_MASK software,
but was better at flagging more rogue pixels so that we did not have
to flag any pixels by hand.
Rejected pixels were interpolated from the surrounding pixels.

The cleaned spectra
were coadded and extracted using the Spectroscopy Modeling
Analysis and Reduction Tool \citep[SMART;][]{Higdon2004}.  
The method of sky subtraction depended on the resolution of the data. 
For the low-resolution
data, sky subtraction was performed by subtracting one nod position
from the other before extraction.  For high-resolution data, the
narrow width of the slit required that separate sky images be
subtracted from the QSO images in each nod position. A separate set of
images 130 arcseconds away from the QSO was taken for this purpose.
This same procedure applied to a standard star (HR 7341) yielded a spectrum
which matched between each of the IRS orders/modules, matched the calibration
spectrum within 5\%, and gave a
spectrum which obeyed the Rayleigh-Jeans limit, so we are confident
of the relative calibration of our data, but expect that the absolute
calibration has an uncertainty of 5\%.  We re-reduced the data
with later versions of the pipeline which resulted in fluxes
that differ by as much as 20\% in the overlapping region between
different orders, while the 14.0.0 pipeline did not have this
problem.

For each order of each module, we fit a Gaussian to the distribution of 
the difference in flux between the two nods divided by the sum of the squares 
of the uncertainties; in all cases the standard deviation of this distribution
differed from unity indicating that the uncertainties were misestimated.  We 
scaled the SMART uncertainties by the standard deviation of this Gaussian.
We then fit the median-smoothed spectrum from all nods and orders with
a 5-th order polynomial, and cleaned the data of points which disagreed by 
$>3 \sigma$ from this fit, as well as points for which the two nods disagreed 
by $>3 \sigma$.  This procedure
automatically removed data near the edges of each order, which are notoriously
unreliable, and also removed other outliers which may be due to improperly
cleaned cosmic rays or rogue pixels.

\subsection{IRAC}

\begin{figure}
\centerline{\psfig{figure=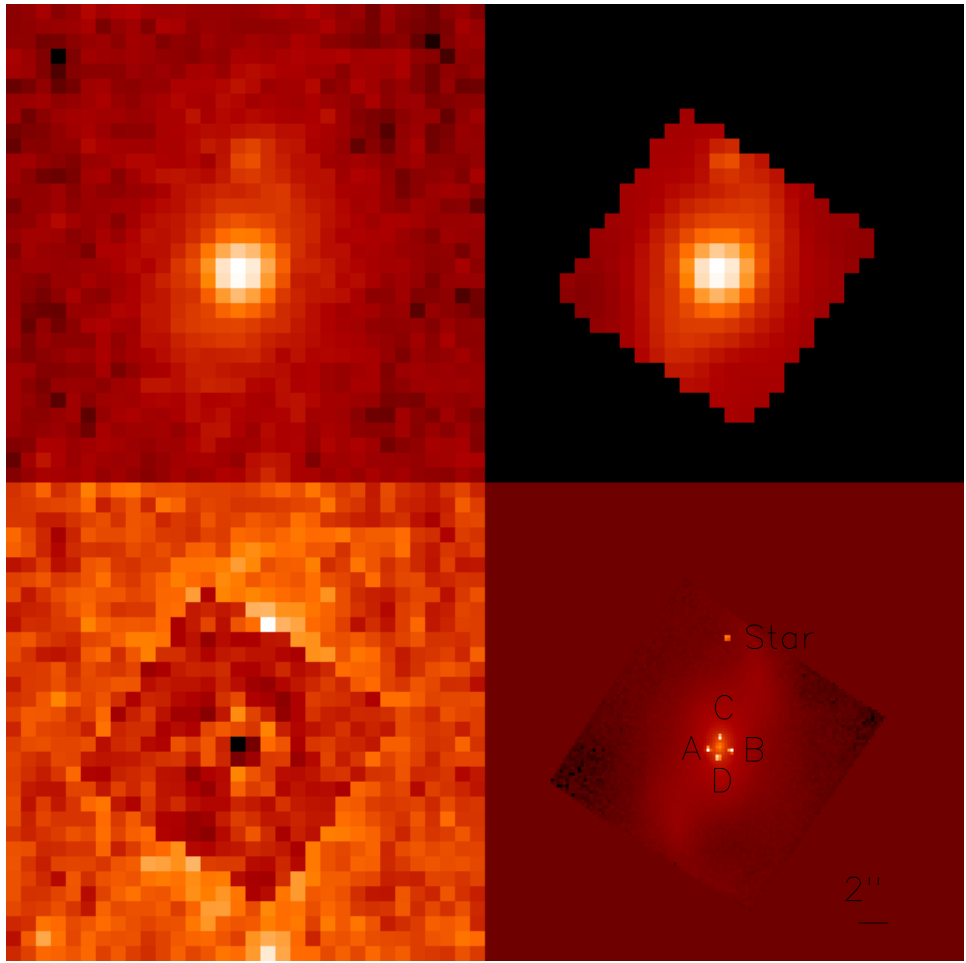,width=5in}}
\caption{{\bf Top left panel}: Cutout of 32$\times$32 pixel region centered on the Channel
2 (4.5 $\mu$m) mosaic, logarithmic intensity scaling.
{\bf Top right panel}: best-fit model to Channel 2 data, including four quasar
images, galaxy scaled from HST H-band, and star. {\bf Lower left panel}: difference
between data and model.  The HST H-band image limits the size of the
region in the model to the central 245 pixels. {\bf Lower right panel}: ``Deconvolved" model 
image at 5 times the resolution of Spitzer with the four QSO images and nearby star labelled.}
\label{fig:irac_aper}
\end{figure}

\qso\ was observed for 38.4 sec in each of the four IRAC wavebands (3.6, 
4.5, 5.8, and 8.0 $\mu$m) in full array mode with a 16-position spiral 
dither pattern with 2-second exposures at each position.  Post-BCD mosaics 
were obtained from the \emph{Spitzer} archive (pipeline version 13.0.2), which
we used in our analysis.
As \qso\ is only a few pixels across at the resolution of IRAC (see
Figure~\ref{fig:irac_aper}), the four lensed images and lens galaxy are 
unresolved; however, we were still able to derive the flux ratios of
the four images.

\begin{figure}
\centerline{\psfig{figure=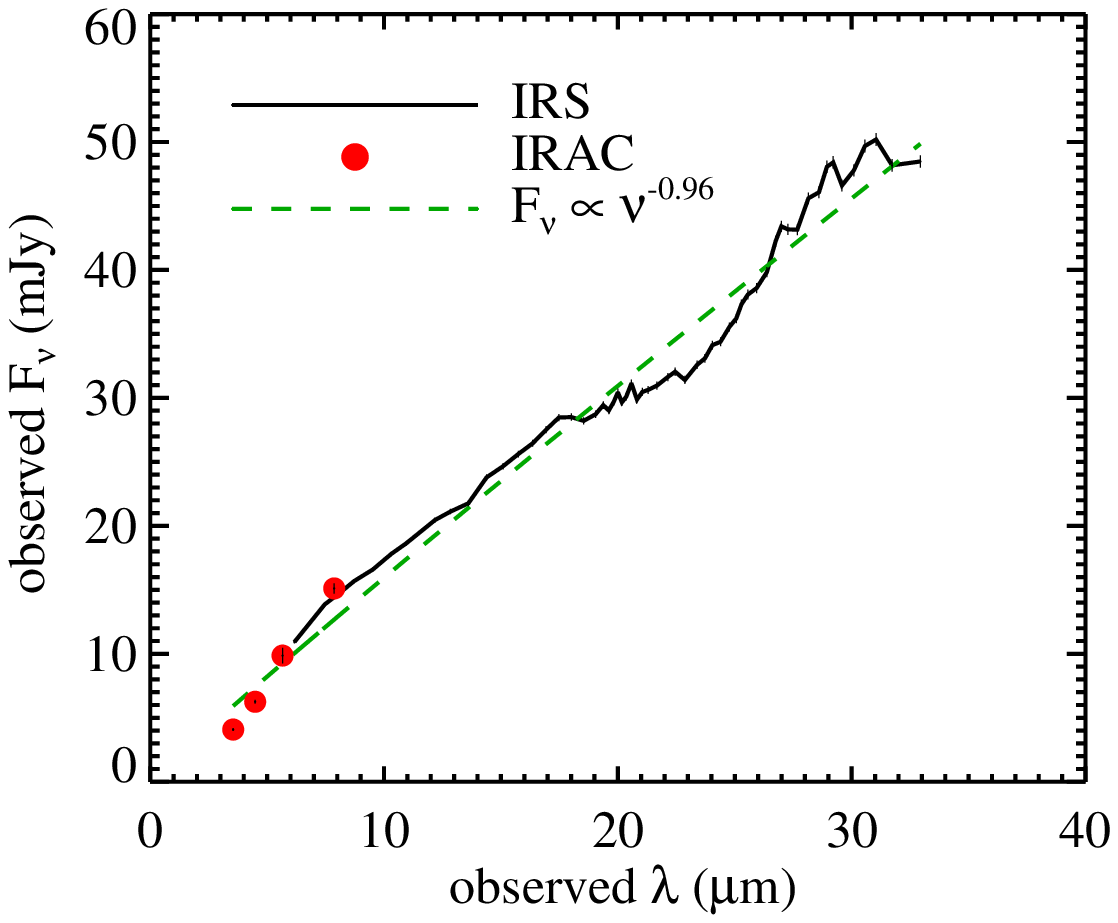,width=5in}}
\caption{Spectrum of \qso.  Black solid line is the IRS spectrum; 
red filled circles are the IRAC photometry; and green dashed line is 
the power-law fit to the entire observed spectrum.}
\label{fig:q2237spect}
\end{figure}

\subsection{Flux ratios}

For comparison with microlensing models, we derived the fluxes of the lensed 
quasar images from the IRAC data.  As the pixel size of the IRAC images, 
$1\farcs2$, is comparable to the separation of the quasar images,
this required a multi-component model fit.  Since the relative positions of the quasar 
images are known extremely accurately, and since the IRAC point spread
function (PSF) is known fairly precisely, we derived the flux ratios 
of the four lensed images of the quasar with PSF fitting.  The main 
uncertainty in the fitting is the contribution of the
lens galaxy to the flux in the IRAC bands; we addressed this by using
the HST H-band as a model from the {\it CASTLES} survey \citep{Kochanek2009}, 
assuming no color gradients between the H-band
and IRAC bands. This is likely a good approximation as extinction and 
intrinsic colors should vary weakly in the infrared since the stellar emission 
is well into the Rayleigh-Jeans tail.

We created a model composed of (i) four quasar images; 
(ii) the one star 10$^{\prime\prime}$ from the center of the lens galaxy; 
(iii) the lens galaxy flux, scaled from a deconvolved HST $H$-band image; 
(iv) a uniform sky background.  
We held the relative positions of the quasar images (and the nearby star) 
fixed to the values derived from the HST H-band image, while we allowed the 
absolute position to vary (given the uncertain absolute pointing of Spitzer).

The lens galaxy was isolated in the HST H-band image by masking the quasar 
images and stars 
(within a circle 14.25 pixels from the location of each point source), 
and the masked region for each quasar image was replaced with an
elliptical Sersic model fit to the remaining H-band data, while the masked 
region near the star was filled in with 
the median flux near its location.  As the Spitzer IRAC PSF is derived
at 5 times the pixel resolution ($0\farcs 24$), we rotated and compressed 
the HST image
to fit the Spitzer images at 5 times the resolution.  We then convolved the 
HST image with the IRAC PSF for each band, multiplying by a constant factor to scale
to each IRAC band, and added to this the five point fluxes by interpolating 
the Spitzer IRAC PSF to the location of each point source and multiplying
by their respective fluxes.  Finally we added in a constant flux to represent 
the sky.

These model components give a total of nine free parameters to fit: 5 point 
sources, the extended galaxy flux scaling factor, the sky flux, and 
the RA and DEC of image A, which was taken as the reference point.  We 
computed the $\chi^2$ of this model by comparing with the Post-BCD mosaic 
and uncertainties from the Spitzer IRAC pipeline.  We optimized the model 
parameters using the Levenberg-Marquardt method, and then found the 
uncertainties on each parameter from a Markov chain computed using the method 
described in \citet{Tegmark2004}.  The best-fit $\chi^2$ for the four IRAC 
channels was (314,209,91,162) for 236 degrees of freedom (9 model parameters 
to fit the flux of 245 pixels which is the region covered by the HST H-band image).  
Formally these fits range from very good to poor, which may indicate that the 
model is inadequate (e.\ g.\ possibly the galaxy has color
gradients between 2.2 and 3.6 microns), or the error bars are misestimated.
We also computed error bars on the model parameters using 
the covariance matrix evaluated at the best fit and by finding the region 
with $\Delta \chi^2 < 1$ for each parameter while marginalizing over the 
other parameters; each of these techniques gave error bars nearly identical 
to the Markov chain.  We converted these values to fluxes in mJy, as well 
as flux ratios, and report the derived fluxes and errors in Table 
\ref{tabflux}.  For the galaxy we report the entire model flux within 5 
pixels (6$^{\prime\prime}$) of the center of the galaxy except for the 
contribution from quasar images.  The $V$-band data is from 
data taken by the {\it OGLE} collaboration one day before the {\it Spitzer}
observations, and the errors reported are relative flux errors, not absolute
\citep{Udalski2006}. In addition we report the continuum spectral slope, 
$\alpha_\nu$ measured at 5400 \AA\ for $f_\nu \propto \nu^{\alpha_\nu}$, 
for all four images measured 
with a VLT observation on 11 November 2005 \citep{Eigenbrod2008a}.


\begin{table}
\caption{\label{tabflux} Fluxes of quasar images in milli-Janskys and optical spectral slope.}
\centerline{
\begin{tabular}{lcccccc}
\hline
Image & $F_\nu(V)$ & $\alpha_\nu$ & $F_\nu(3.6\mu$m) & $F_\nu(4.5\mu$m) & $F_\nu(5.8 \mu$m) & $F_\nu$(8.0$\mu$m) \\
\hline
A          & $0.507 \pm 0.005$ & $-1.064\pm 0.002$ & $ 1.60 \pm  0.07$ & $ 1.99 \pm  0.08$ & $ 3.21 \pm  0.37$ & $ 4.49 \pm  0.22$ \\
B          & $0.257 \pm 0.004$ & $-0.859\pm 0.004$ & $ 1.14 \pm  0.06$ & $ 1.73 \pm  0.06$ & $ 2.78 \pm  0.28$ & $ 4.19 \pm  0.19$ \\
C          & $0.197 \pm 0.004$ & $-1.374\pm 0.005$ & $ 0.48 \pm  0.05$ & $ 1.25 \pm  0.07$ & $ 1.60 \pm  0.27$ & $ 2.67 \pm  0.16$ \\
D          & $0.185 \pm 0.005$ & $-1.335\pm 0.006$ & $ 0.85 \pm  0.06$ & $ 1.28 \pm  0.08$ & $ 2.27 \pm  0.33$ & $ 3.76 \pm  0.22$ \\
Total A-D  & $ 1.15 \pm  0.01$ &                   & $ 4.08 \pm  0.11$ & $ 6.25 \pm  0.15$ & $ 9.86 \pm  0.60$ & $15.11 \pm  0.40$ \\
Star       &                   &                   & $ 0.29 \pm  0.01$ & $ 0.18 \pm  0.01$ & $ 0.13 \pm  0.05$ & $ 0.00 \pm  0.03$ \\
Galaxy     &                   &                   & $ 5.34 \pm  0.08$ & $ 3.52 \pm  0.09$ & $ 2.87 \pm  0.35$ & $ 1.96 \pm  0.23$ \\
\hline
\end{tabular}}
\end{table}


\begin{table}
\caption{\label{tabratio} Fluxes ratios of quasar images (IRAC)}
\centerline{
\begin{tabular}{lccccc}
\hline
Image & V band & 3.6 $\mu$m & 4.5 $\mu$m & 5.8 $\mu$m & 8.0 $\mu$m \\
\hline
B/A          & $ 0.507 \pm  0.010$ & $ 0.711 \pm  0.061$ & $ 0.869 \pm  0.053$ & $ 0.867 \pm  0.195$ & $ 0.932 \pm  0.073$ \\
C/A          & $ 0.389 \pm  0.009$ & $ 0.298 \pm  0.040$ & $ 0.626 \pm  0.041$ & $ 0.499 \pm  0.099$ & $ 0.593 \pm  0.043$ \\
D/A          & $ 0.366 \pm  0.010$ & $ 0.531 \pm  0.049$ & $ 0.643 \pm  0.054$ & $ 0.709 \pm  0.177$ & $ 0.837 \pm  0.075$ \\
C/B          & $ 0.767 \pm  0.020$ & $ 0.418 \pm  0.051$ & $ 0.720 \pm  0.051$ & $ 0.576 \pm  0.115$ & $ 0.637 \pm  0.043$ \\
D/B          & $ 0.721 \pm  0.023$ & $ 0.746 \pm  0.080$ & $ 0.740 \pm  0.054$ & $ 0.817 \pm  0.133$ & $ 0.898 \pm  0.069$ \\
D/C          & $ 0.940 \pm  0.032$ & $ 1.783 \pm  0.319$ & $ 1.027 \pm  0.104$ & $ 1.420 \pm  0.431$ & $ 1.411 \pm  0.146$ \\
A/(A+B+C+D)  & $ 0.442 \pm  0.018$ & $ 0.394 \pm  0.017$ & $ 0.319 \pm  0.011$ & $ 0.325 \pm  0.036$ & $ 0.297 \pm  0.013$ \\
B/(A+B+C+D)  & $ 0.224 \pm  0.009$ & $ 0.280 \pm  0.014$ & $ 0.277 \pm  0.010$ & $ 0.282 \pm  0.029$ & $ 0.277 \pm  0.011$ \\
C/(A+B+C+D)  & $ 0.172 \pm  0.008$ & $ 0.117 \pm  0.014$ & $ 0.199 \pm  0.010$ & $ 0.162 \pm  0.026$ & $ 0.176 \pm  0.010$ \\
D/(A+B+C+D)  & $ 0.162 \pm  0.008$ & $ 0.209 \pm  0.016$ & $ 0.205 \pm  0.013$ & $ 0.230 \pm  0.034$ & $ 0.249 \pm  0.015$ \\
\hline
\end{tabular}}
\end{table}

\section{Results}

\subsection{Spectral Energy Distribution} \label{sedsection}

In this section we compare the \emph{Spitzer} spectral energy distribution of
\qso\ to Seyfert galaxies and quasars to show that it looks like
a typical radio-quiet active galaxy in the infrared.
Figure~\ref{fig:q2237spect} shows the \emph{Spitzer} spectrum of
\qso\ which has been binned so that each bin has a signal-to-noise
greater than 100.  The excellent agreement between the IRS and IRAC 
results, which had completely independent flux calibration, bolsters 
our confidence in the accuracy of our reported fluxes.

We fit a power law of the form $F_{\nu} \propto \nu^{\alpha}$ to the
IRS spectrum, and we find $\alpha = -0.96 \pm .02$, giving a spectrum which
is nearly flat in $\nu F_\nu$.

\begin{figure}
\centerline{\psfig{figure=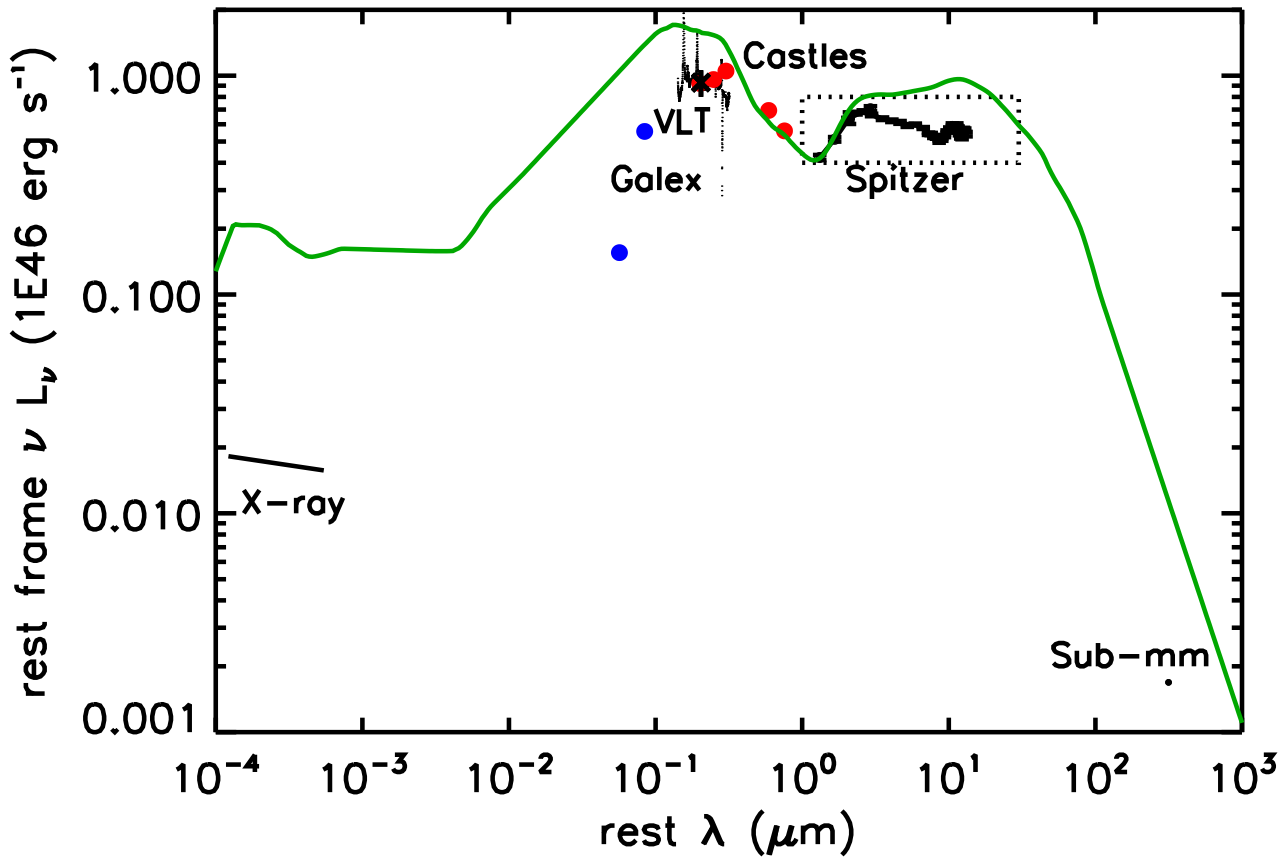,width=\hsize}}
\caption{Full isotropic spectral energy distribution of \qso.  Black
line is X-ray data from {\it Chandra}, blue points are from {\it GALEX},
black spectrum near the peak (big-blue bump) is from the VLT,
red points are from the {\it Castles} database and from {\it OGLE}, 
black infrared
points are {\it Spitzer} data from this paper, and the black point
in the lower right is from ground-based sub-mm observations;
green line is quasar composite spectrum from \citet{Elvis1994}
normalized to the 1$\mu$m dip. The dotted box shows the region
plotted in Figure \ref{fig:q2237_netzer}.}
\label{fig:models}
\end{figure}

Figure~\ref{fig:models} shows the full spectrum of \qso; the
isotropic luminosity is defined as $(\nu L_\nu)_{rest} = 4\pi D_L^2 (\nu F_\nu)_{obs} /\mu$, where $rest$/$obs$ refer to the rest-frame/observed
frequencies, $D_L$ is the luminosity distance of the quasar,
and $\mu$ is the total magnification of the quasar.  Included are
our data from both IRS and IRAC, optical and near-infrared data points from
the CfA-Arizona Space Telescope LEns Survey \citep{Falco2001}, 
OGLE \citep{Udalski2006}, and
\citet{Eigenbrod2008a}, X-ray data from \citet{Dai2003},
two data points from the GALEX archive \citep{Martin2005},
and a sub-mm data point at 850 $\mu$m from \citet{Barvainis2002}.  To
compute the total luminosity we have assumed the cosmological parameters
from the WMAP 5-year data set \citep{Dunkley2008} as well as a total macrolensing
magnification of $\mu=16$ \citep{Schmidt1998}.  The optical
data have been corrected for extinction in the Milky Way assuming a Galactic reddening
of $E(B-V)=0.068$ with $R_V=3.1$ extinction curve.  As the light from the quasar
passes through different portions of the bulge of the lens galaxy, we need to make 
additional extinction corrections for the four lensed images.
We have used the flux ratios of the broad lines averaged over time from 
the data set of \citet{Eigenbrod2008a} to derive the {\it relative} extinction of
the four quasar lensed images.  We find image B is extincted relative to image
A by $\Delta E(B-V) = 0.02 \pm 0.05$, while images C and D are reddened 
with respect to images A and B by $\Delta E(B-V)= 0.10 \pm 0.04, 0.18 \pm 0.03$, respectively.
Since images A and B have small (possibly zero) relative extinction, we assume
that each of these images has zero extinction in the lens galaxy, and simply
correct images C and D.  We have not attempted to correct the data
for microlensing, so the overall uncertainty is at least
0.2 mag.   From the \qso\ SED we find a total luminosity of $L_{tot}=4.0 
\times 10^{46}$ erg/s.

\begin{figure}
\centerline{\psfig{figure=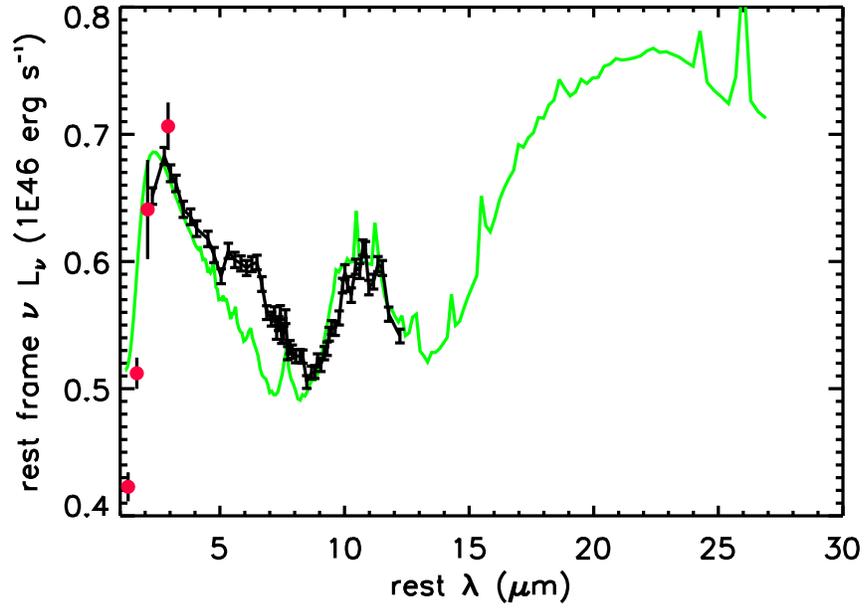,width=5in}}
\caption{Comparison of the \qso\ SED with the composite infrared spectrum
for far-IR weak quasars from \citet{Netzer2007} (light green solid curve).  
Red filled circles are IRAC data; black connected points with error bars
are binned IRS data.} 
\label{fig:q2237_netzer}
\end{figure}

In Figure~\ref{fig:models} we plot the SED of \qso, and for reference
compare it to the composite radio-quiet 
quasar SED from \citet{Elvis1994}, normalized to 1.3 $\mu$m.   Although
\qso\ appears underluminous in the X-ray, UV, mid-IR, and sub-mm relative
to the composite, this behavior is well within the range of SEDs in the Elvis sample, and it is likely that the composite is affected by selection biases
at these wavelengths where many quasars were not detected.  The SED of \qso, 
shown in  Figure \ref{fig:q2237_netzer}, looks fairly typical compared to  a 
composite spectrum of Palomar-Green quasars with weak far-infrared 
emission \citep{Netzer2007}.  There are minor differences such as an extra bump 
near 6-7 $\mu$m and a peak associated with the hottest dust at slightly longer 
wavelengths ($\sim 2.8\pm 0.3 \mu$m), but these differences are well within the 
range of variation within the PG quasar sample.  If we ``fit" the \qso\ SED by 
scaling the Netzer composite spectrum by an arbitrary factor, we find a $\chi^2 
= 573$ for 32 DOF, which is formally a very poor fit, but the discrepancy 
is dominated 
by the disagreement in the cutoff at short wavelengths and the bump near 7 microns.

\begin{figure}
\centerline{\psfig{figure=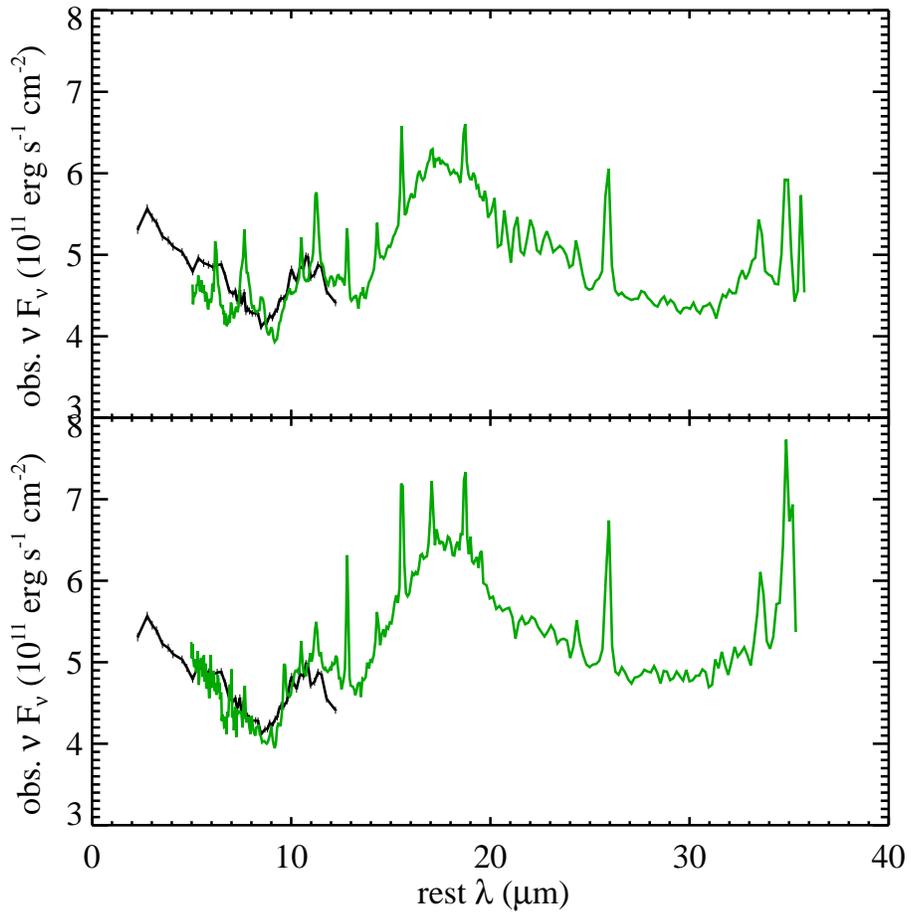,width=5in}}
\caption{Black connected points with error bars: IRS spectrum of \qso.
Green solid lines: spectra of Seyfert galaxies Mrk509 (top) and
MCG-2-58-22 (bottom) scaled to match flux of \qso.}
\label{fig:seyferts}
\end{figure}

The infrared spectrum of \qso\ also looks very similar to
low redshift Seyfert galaxies taken from a sample of 
23 galaxies (Gorjian et al., in preparation).  The
\emph{Spitzer} spectrum of \qso\ is plotted with the two
most similar Seyfert spectra in Figure~\ref{fig:seyferts}, scaled 
to match the flux of \qso.  The qualitative shape of
the SEDs matches well from 4-10 $\mu$m, although Mrk 509 shows
stronger emission features, presumably due to silicates.

The similarity of the infrared SED of \qso\ to other Seyferts and 
quasars indicates that our microlensing studies of this object will 
broadly apply to radio-quiet active galaxies.

\subsection{Dust emission model}

To improve our physical understanding of the emission from \qso\,
we fitted the SED with the models of \citet{Fritz2006}.
The models utilize the Mathis-Rumpl-Nordsieck (MRN) dust 
size distribution \citep{Mathis1977} with scattering and absorption
opacities from \citet{Laor1993}.  The model fixes the geometry as a torus
with an opening angle that is independent of radius and an inner radius of
the torus,
\begin{equation}
R_{min} = 1.3 {\rm pc} L_{46}^{1/2} T_{1500}^{-2.8},
\end{equation}
which is set by a dust sublimation temperature of 1500 K,
where $L_{46}$ is the AGN luminosity in units of $10^{46}$ erg s$^{-1}$.
The dust density in the model is described by $\rho(r,\theta) \propto
r^\beta e^{-\gamma |\cos{(\theta)}|}$, with a dust-free
cone within polar-angle $\theta < \theta_{cone}$.
The grid of models covers a range of parameters for the dust with: (1)
the ratio of the outer to inner radius, $30 < R_{max}/R_{min} < 300$;
(2) the variation of dust density with radius, $-1 < \beta < 1/2$;
(3) the equatorial optical depth at 9.7 $\mu$m, $0.1 < \tau_{9.7} < 10$;
(4) inclination angles, $i$, ranging from 0.01 (edge-on) to 89 degrees (face-on),
and $11,21,31,41,51,61,71,81^\circ$ in between;
(5) dust-free cone with size $20^\circ < \theta_{cone} < 60^\circ$;
and (6) an angular cutoff in density with $0 < \gamma < 6$.

\begin{figure}
\centerline{\psfig{figure=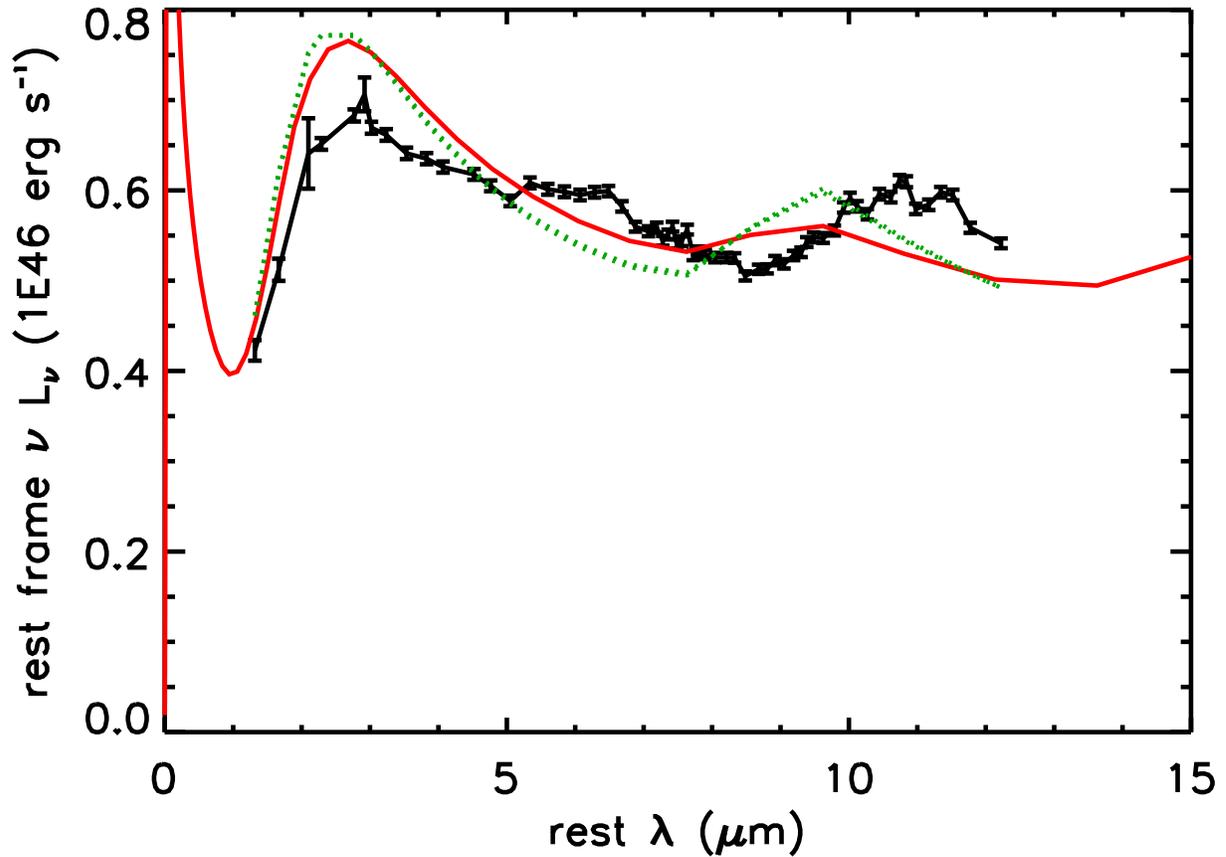,width=\hsize}}
\caption{Best-fit model from \citet{Fritz2006} plotted versus
the Spitzer rest-frame spectrum of \qso\, as described in
the text.  Data is solid black line connecting points with error
bars; best-fit model (71$^\circ$) is solid red line, while dotted green lines
(which appear as one line) are the same model viewed at 81$^\circ$ 
and 89$^\circ$.}
\label{fig:q2237_fritz}
\end{figure}

We have scaled the infrared portion of the models by an arbitrary
constant to optimize the match with our Spitzer SED.  The  best-fit
model is shown in Figure \ref{fig:q2237_fritz}, which has parameters 
$R_{max}/R_{min} = 100$, $\beta = -1/2$, $\tau_{9.7} = 10$, 
$i = 71^\circ$, $\theta_{cone} = 20^\circ$, and $\gamma=3$.  The
best fit implies an AGN luminosity of $4.4 \times 10^{46}$ erg s$^{-1}$,
remarkably close to the luminosity found from the \qso\ SED 
($L=4.0 \times 10^{46} erg/s$), despite
the fact that we have only fit the Fritz models to the infrared data.
This may imply that the dust acts as a fairly good calorimeter of the
total AGN flux.
The qualitative fit to the data is fair: the model shows a peak
near 2.5 $\mu$m in the rest frame, and gradual decline towards
longer wavelengths, and an emission feature near 10$\mu$m.  However,
quantitatively the fit is horrible: $\chi^2= 3045$ for 94 degrees of
freedom.  This is primarily due to the fact that the short-wavelength
peak is more prominent in the model than in the observations and the silicate 
absorption and emission features in the model are offset in wavelength
of the observed features.  It is possible that optimizing the parameters
will improve the fit as the grid is quite coarse and some of
the best-fit parameters are at the extreme values of the grid,
such as $\theta_{cone}$.  Also, the viewing angle is $71^\circ$ which
is only 1$^\circ$ within the opening angle of the cone; however, viewing
angles of 81$^\circ$ and 89$^\circ$ are very similar in shape, but only
slightly poorer fits, plotted as dotted lines in Figure \ref{fig:q2237_fritz}.
Consequently we do not believe that the fitted parameters are unique
or even correct; indeed the simple geometry chosen by Fritz et al.\ may be
wrong.  The main point is that a dusty torus model can produce
a fair qualitative fit to the SED of \qso; further development of
theoretical models will be required to obtain a better quantitative fit.

\subsection{Measured flux ratios}

\begin{figure}
\centerline{\psfig{figure=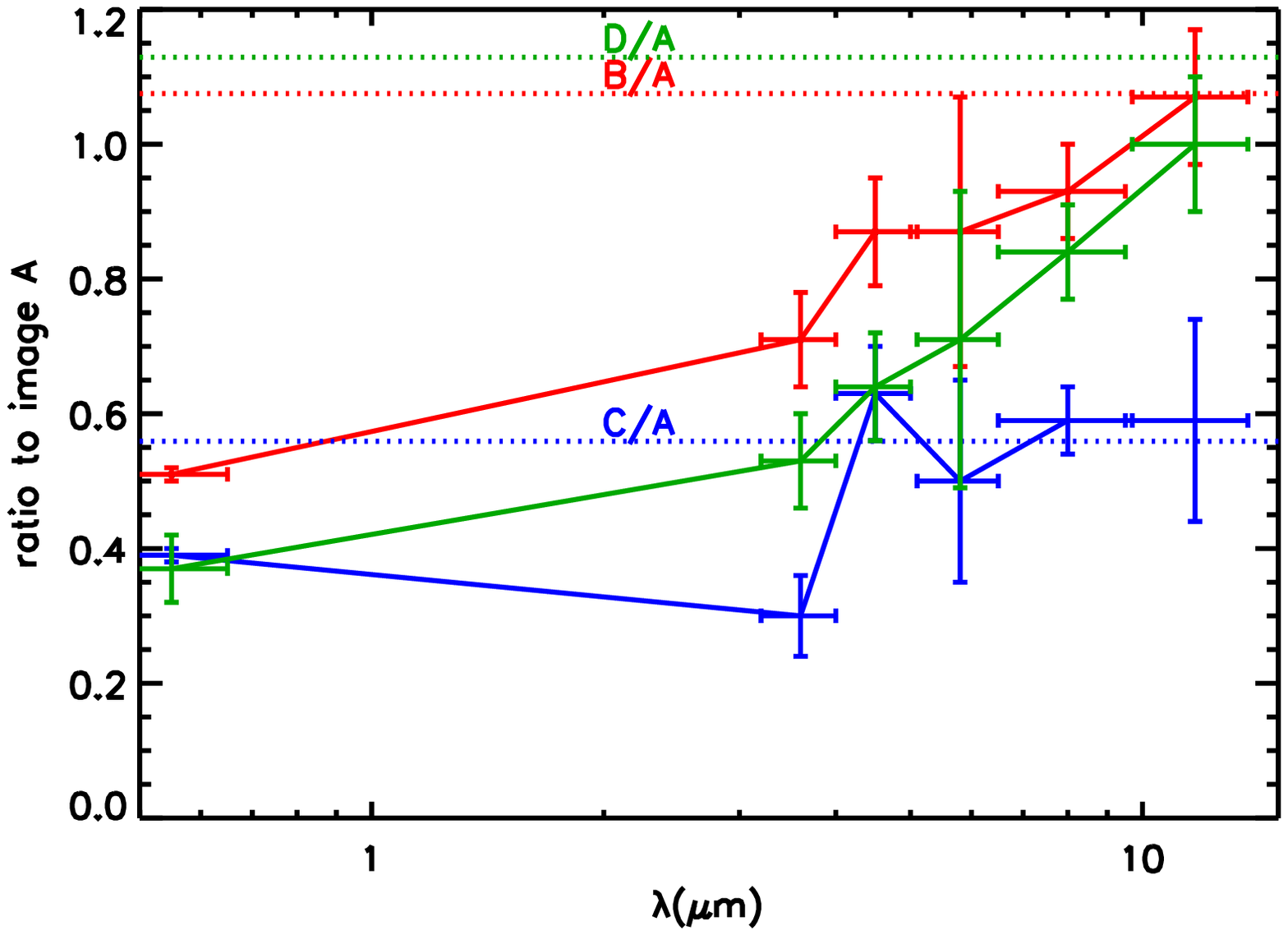,width=\hsize}}
\caption{Flux ratios of \qso\ as a function of wavelength.  Red points 
show flux ratios of image B to image A, blue are ratio of image C to image
A, and green are ratios of image D to image A.  The V-band data (left-most
points) are
taken from the OGLE website at the time closest to the IRAC observations;
the 11 micron data are taken from \citet{Agol2000}.  The dotted
lines are the flux ratios predicted from the model of \citet{Trott2002}.}
\label{fig:q2237ratios}
\end{figure}

Figure \ref{fig:q2237ratios} shows the ratios of the quasar image fluxes as
a function of wavelength from Table \ref{tabflux}.  The $V$-band data are 
the data obtained from the OGLE data archive \citep{Udalski2006} taken at a 
time closest to our IRAC observations: 2005 Nov.\ 16 00:48 UT (HJD).
The 10 micron points are from \citet{Agol2000}, and thus are not
simultaneous to our Spitzer observations; however there appears to be 
little variability at this wavelength.   Also plotted are the model flux
ratios from
\citet{Trott2002} which is the most complete model of the lens galaxy
of \qso\ constructed to date, including
the bar and spiral arms. The agreement with the 10 micron flux
ratios is expected since these were used as a constraint on the model; 
however, the model gives very similar flux ratios as an earlier model by
\citet{Schmidt1998} which was constructed before the 10 micron observations.

The general trend is obvious in Figure \ref{fig:q2237ratios}: for all three
pairs of images there is a strong microlensing anomaly in the optical which 
gradually disappears
towards longer wavelengths.  This is precisely the behavior expected
from the standard model of quasars:  the source should be larger at longer 
wavelengths and thus less affected by microlensing. 

\subsection{Two-component model of flux ratios}

Ideally we would like to use the microlensing flux ratios to constrain
the size of the quasar as a function of wavelength; however, only
weak constraints can be derived without a detection of time variability
due to microlensing \citep{Wyithe2002}.  Instead we use
a semi-empirical model for the infrared SED to {\it predict} what
the flux ratios should be as a function of wavelength and compare
these predictions with the observed flux ratios to confirm the
plausibility of this model.

\begin{figure}
\centerline{\psfig{figure=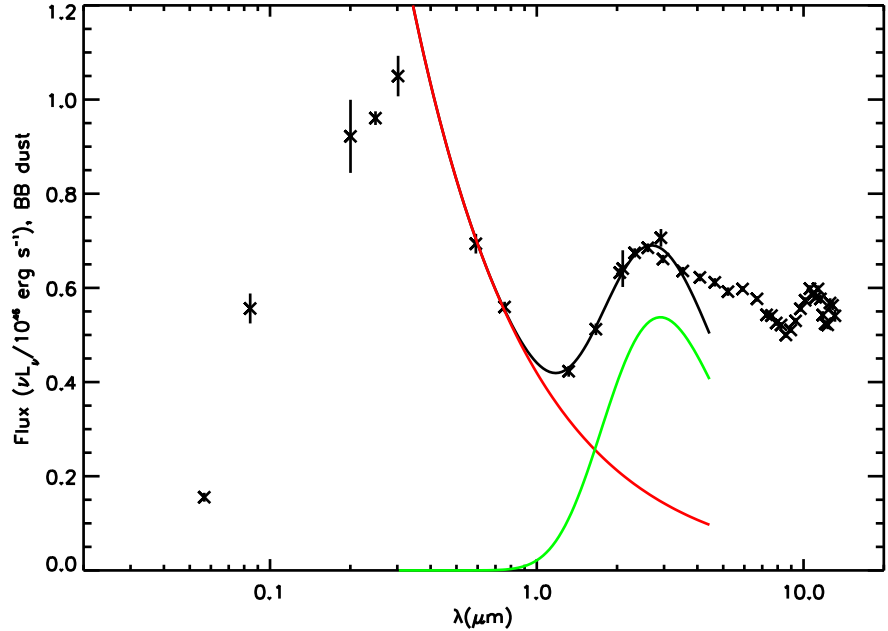,width=5in}}
\caption{Rest-frame spectral energy distribution of \qso\ fit from 
0.4-4 $\mu$m with a power-law (red solid line) and thermal dust emission (green solid line) component. The black solid line is the best-fit sum of both 
components.}
\label{fig:q2237sed}
\end{figure}

We constructed a semi-empirical model for the flux ratios to compare
to the observed IRAC flux ratios as follows.  We assume that the spectrum
consists of a power-law component due to an accretion disk (we
are modeling the region from $0.4 - 4.0 \mu$m which is well longward
of the peak of the disk spectrum and is only a decade
in frequency, so a power-law should be an adequate approximation of
a disk spectrum)
and a single-temperature thermal dust emission component, representing
the inner edge of the a dusty torus.  We fit the spectral energy 
distribution
from 0.4 to 4.0 microns in the rest frame with these two components,
determining their relative strength at each wavelength.  The
best fit is shown in Figure \ref{fig:q2237sed}; the model provides
a good fit to the four IRAC data points.  We have not attempted to correct for 
microlensing, nor possible time-variability as the SED data are
not simultaneous.  However, this will likely have a small effect on
the SED as summing over all four images reduces the impact of
microlensing and in the infrared quasars are weakly variable.

With these two fits we determined the minimum possible source sizes
to reproduce the observed flux with thermal emission as follows.  For the
power-law component, we assumed a disk geometry with a temperature
that is a power-law in radius, $r$, finding $T \propto r^{-0.66}$, and 
found that the half-light radius should scale with wavelength as
\begin{equation} \label{sourcesize}
r_{1/2} = 4 \times 10^{16} {\rm cm} \lambda^{1.5},
\end{equation}
where $\lambda$ is measured in microns in the rest frame of the quasar.
At this radius the standard disk model is well outside the inner edge
and thus is expected to have a temperature
dependence of $T \propto r^{-3/4}$, which is close to the measured dependence.
We assumed that the dust component either has an emissivity described
by optically-thin interstellar medium (ISM) dust with the model of 
\citet{Draine2003} or emits as an optically-thick blackbody (BB).
These two extremes were chosen to bracket the range of possible behaviors
for the hottest dust at the inner edge of the torus (we did not use
the best-fit Fritz model due to the different peak wavelength).
We found best-fit temperatures of the dust of 1168 K (ISM) or 1264 K
(BB). This requires a {\em minimum} distance from the quasar of 3.83 
pc (ISM) or 0.76 pc (BB) to maintain temperature equilibrium, assuming
that the quasar emits isotropically.
The total luminosity in this component is $6 \times 10^{45}$ erg/s
(ISM/BB) implying an emission region of at least 1.3 pc (ISM)
or 2.2 pc (BB)  in size, consistent with the radiation equilibrium 
argument within a factor of 3.

If the average mass of microlenses is 0.1 $M_\odot$, then the
Einstein radius is $r_E = 5.77 \times 10^{16}$~cm projected to
the source plane.  At one micron the power-law emission (if
thermal) is comparable in size to an Einstein radius, while the
dust emission component is about 200 times larger than the
Einstein radius.  Thus, the power-law component will be affected
by microlensing, while the dust component should be nearly unaffected
by microlensing, with variations of less than 1\% \citep{Refsdal1991}.

To model the microlensing of the power-law component, we
extrapolate the wavelength dependence of the optical flux ratios
measured with the VLT \citep{Eigenbrod2008a} to the infrared.
We correct the optical flux ratios for differential extinction
using the E(B-V) values derived from broad emission lines, as
described in section \S \ref{sedsection}.  Then, the expected
flux ratio as a function of wavelength, $r_{BA}(\lambda)$, of 
images A and B is given by:
\begin{equation}
r_{BA}= {\mu^B_{P} F_{P} + \mu^B_{D} F_{D}
\over \mu^A_{P} F_{P} + \mu^A_{D} F_{D}} e^{-\tau_B+\tau_A},
\end{equation}
where $\mu^{A,B}_{P,D}(\lambda)$ are the magnifications of the
power-law ($P$) and dust ($D$) components as a function of wavelength,
$F_{P,D}(\lambda)$ are the intrinsic (un-magnified) fluxes of these
two components, and $\tau_{A,B}(\lambda)$ are the optical depths
through the lens galaxy for each component (we have dropped $\lambda$
in this equation for simplicity).

Now, as argued above, the dust component should be large enough to
be unaffected by microlensing, so $r_{BA,D}=\mu^B_{D}/\mu^A_{D}$ can
be derived from a model for microlensing.  We utilize the model of
\citet{Schmidt1998}, improved upon by \citet{Trott2002},
with more recent modifications based on kinematic 
data \citep{Trott2008}.  The relative strengths of
the power-law and dust components we take from the model of the
spectral energy distribution (Figure \ref{fig:q2237sed}), $f_{DP}=F_{D}(\lambda)/
F_{P}(\lambda)$; as mentioned above; as the SED
is constructed from the sum of all four images it should be
affected only weakly by microlensing.  Finally, the wavelength
dependence of the microlensing magnification we take from the
extinction-corrected optical flux ratios measured with the
VLT, $r_{BA,P} = \mu^B_{P}/\mu^A_{P}$.  For the extinction correction we
assume a Milky Way extinction curve with $R=3.1$.  Dividing through
the numerator and denominator by $\mu^A_{P} F_{P}$,
we can rewrite the above equation as 
\begin{equation} \label{ratioequation}
r_{BA} = { r_{BA,P} + (\mu^A_{D}/\mu^A_{P}) r_{BA,D} f_{DP}
\over 1 + (\mu^A_{D}/\mu^A_{P}) f_{DP}} e^{-\tau_B + \tau_A}.
\end{equation}
The remaining unknown in this equation is $\mu^A_{D}/\mu^A_P(\lambda)$
since the microlensing magnification of the power law component for each
image is unconstrained by our data.  Fortunately the left hand side of this
equation is weakly dependent on this ratio.  We use the size distribution
derived from the power-law emission spectrum (equation \ref{sourcesize})
to compute the probability distribution as a function of wavelength
from microlensing simulations of each image using the macrolensing parameters
from the model of \citet{Schmidt1998}.  We utilized the code of 
\citet{Wambsganss1999} to run the simulations, creating 10 simulations for
each image of a size $20 r_E \times 20r_E$, and then convolving the
magnification pattern with a Gaussian source with the same half-light
radius as derived in equation \ref{sourcesize}.

\begin{figure}
\centerline{\psfig{figure=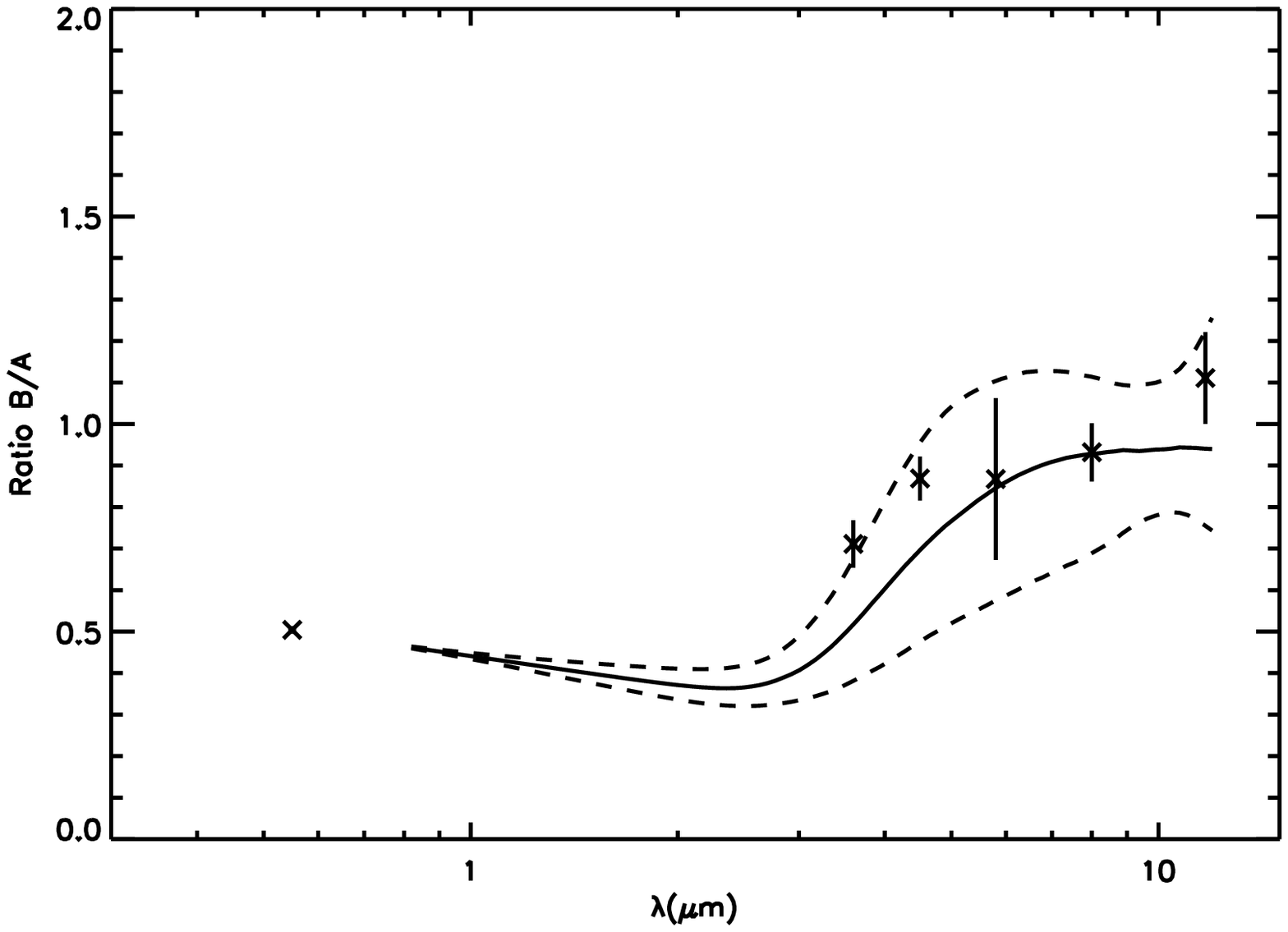,width=3.75in}
\psfig{figure=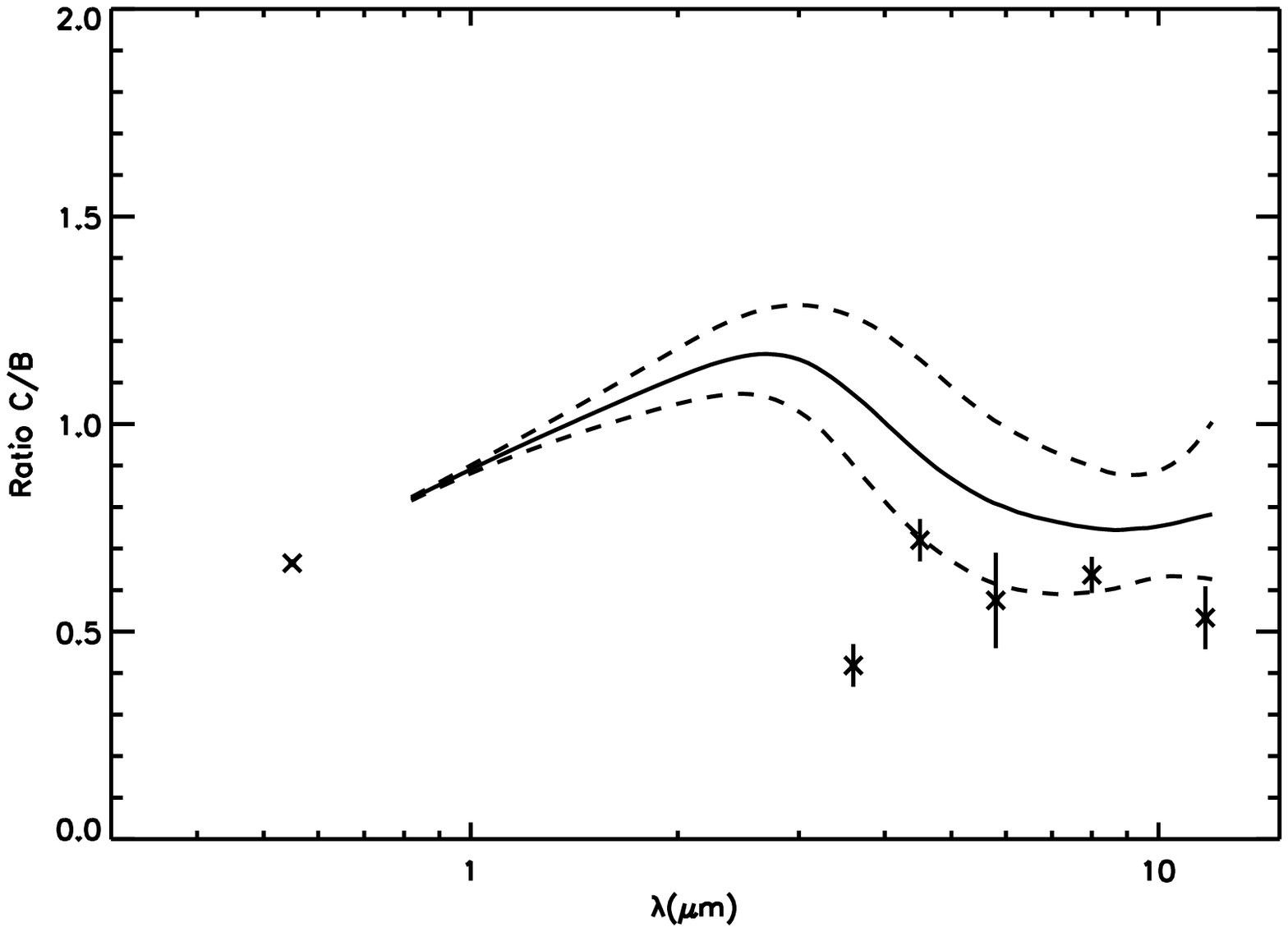,width=3.75in}}
\centerline{\psfig{figure=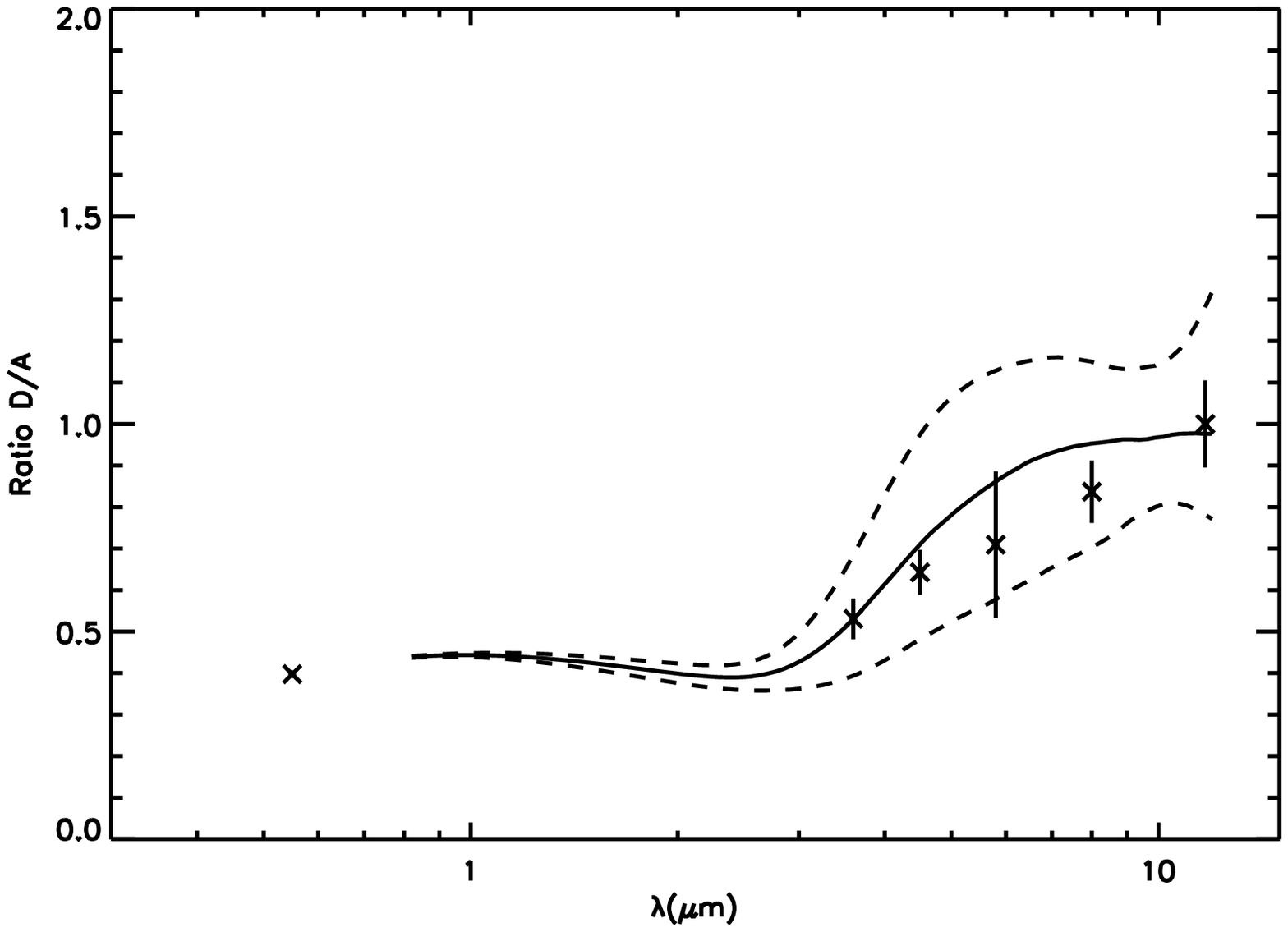,width=3.75in}
\psfig{figure=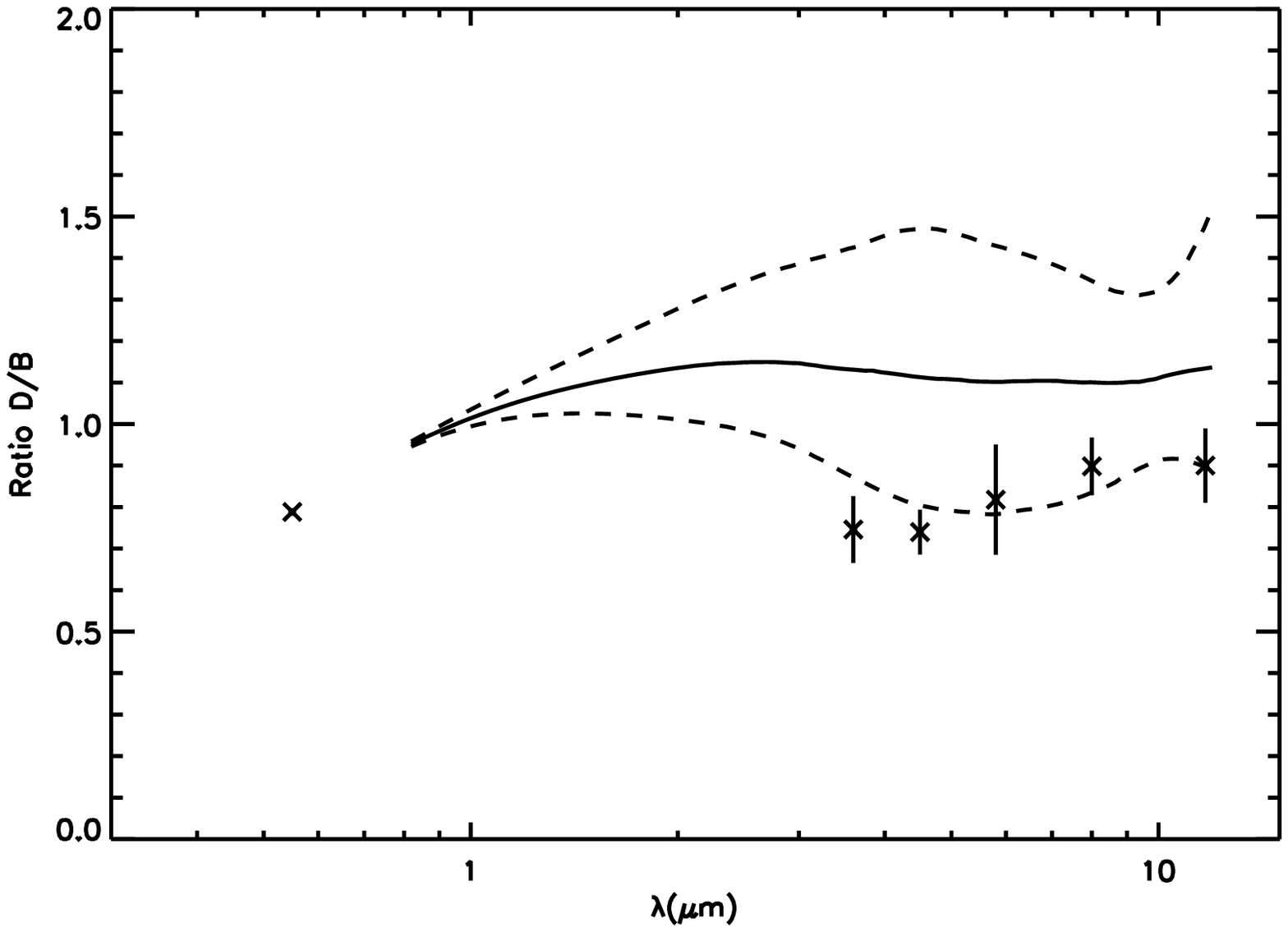,width=3.75in}}
\caption{Flux ratios of images B/A, C/B, D/A and D/B versus 
observed wavelength (crosses), median (solid) and 68.3 percentile (dashed)
prediction of microlensing model of accretion disk and dusty torus.} 
\label{fig:q2237fluxratios}
\end{figure}

To predict the flux ratios at IRAC wavelengths, we have run $10^4$
Monte Carlo simulations sampling the relative extinction, the
optical flux ratios, the galaxy macro-lensing model flux ratios,
and the relative microlensing magnification for the two images 
within the uncertainties of each parameter.  We have then computed
the median and 68\% confidence limits at each wavelength from
these simulations, which is plotted in Figure \ref{fig:q2237fluxratios},
for the ratio of images B to A, as well as several other image pairs.
We find good agreement between the model predictions and the observed
flux ratios.  For all but 3 flux ratios the data are within the 68\%
confidence limits of the model flux ratios.  Thus the wavelength
dependence of the flux ratios is consistent with the
accretion disk/dusty torus model.

\section{Discussion}

The primary two results in this paper are: (1) a measurement
of the infrared spectral energy distribution of the Einstein
Cross; (2) a derivation of the infrared flux ratios in the
mid-infrared (3.6, 4.5, 5.8 and 8.0 $\mu$m observed;
1.3, 1.7, 2.2 and 3.0 $\mu$m in the rest frame) and comparison
to a microlensing model.  We discuss the implications of
these results in this section.

\subsection{Comparison with prior work}

\citet{Agol2000} observed of \qso\ with the Long 
Wavelength Spectrometer on the Keck I telescope at 8.9 and 11.7
$\mu$m;  the flux ratios at these two wavelengths were identical 
within the errors, and both agreed with the macrolensing flux
ratios as predicted by the best lens models.  The flux at 11.7 
$\mu$m reported in that
paper agrees well with our IRS spectrum; however, the 8.9$\mu$m
Keck data point is higher than the IRS data by about 40\% indicating
that the Keck data had an incorrect calibration.  We have not
tracked down the source of this discrepancy, but it cannot
be due to microlensing which would have caused a difference in the
flux ratios at 8.9 and 11.7$\mu$m.  We have more confidence
in the calibration of the \emph{Spitzer} spectrum;
consequently the Keck flux at 8.9 $\mu$m was likely in error.
The qualitative agreement between the \emph{Spitzer} SED and
the Netzer composite indicates that \qso\ behaves as a typical
quasar in the near infrared.

\subsection{\qso\ SED}

As mentioned above, the QSO spectral energy distribution,
Figure \ref{fig:q2237_fritz}, is 
qualitatively well fit with an AGN torus model from \citet{Fritz2006};
however, the publicly available models have a fixed temperature for the
inner edge of the torus at 1500 K which is somewhat higher than
the temperature of the inner edge we have estimated (1164-1250 K).
In addition, the silicate spectral features of the model are a poor
fit to the SED, as is commonly seen in comparing dusty torus models
to AGN SEDs \citep[e.g.][]{Nenkova2002}.  We have found that some Type 
II models (obscured
quasar) fit the silicate feature well, having an offset silicate
feature due to radiation transfer effects, while these models 
do not fit shorter wavelengths which are obscured.  
Thus, it may be possible that the dust composition and/or torus 
opening angle changes with radius causing the silicate feature to 
appear more like that of Type II quasars, while allowing the inner edge
and quasar to be visible so that shorter wavelengths look more
like a Type I quasar.  Another possibility is that the silicate
properties are modified near quasars, either due to changes in the
grain size distribution or due to grain porosity, causing the silicate spectral
features to be shifted \citep[e.g.][]{Voshchinnikov2008}.  Both possibilities
need to be explored in future models of dust grain opacity, as well as
computing better physical models for the dusty torus, such as 
\citet{Krolik2007}.

\subsection{Microlensing, flux ratios, and spectral components}

The agreement between the measured and predicted flux ratios is quite
good, close to $1-\sigma$ for all data points except one
(Figure \ref{fig:q2237fluxratios}).
The uncertainties
in the flux ratio predictions are highly correlated between all wavelengths
since microlensing and extinction have a monotonic variation with wavelength,
so the case of the ratio of images C to B (for example) is still highly
probable.  It is clear from Figure \ref{fig:q2237fluxratios} that an
extrapolation of the optical power law (which can be seen shortward of 1
micron) {\it does not} do a good job
of predicting the infrared flux ratios, while including the un-microlensed
infrared bump due to dust emission brings the flux ratios back into 
agreement with the data (within one standard deviation).   If the
power-law component had a cutoff around one micron so that the infrared
data were {\it solely} due to the extended dust emission, then the flux
ratios would change abruptly to the macrolensed values.  This is not
the case for, e.g., the ratio of image D to image A, indicating that
{\it both} the power-law and dust emission components are required to
fit the IRAC data.  This adds evidence to the case for the presence
of an accretion disk emission component under the infrared bump.

By extrapolating the wavelength dependence of the microlensing flux ratios from 
the optical, we mostly avoided needing the size of the accretion disk in
units of the Einstein radius.  The only place the size of the accretion 
disk enters our analysis is in computing the flux ratios of the disk and torus
components for each image in equation \ref{ratioequation} ($\mu^A_D/\mu^A_P$, 
and the same ratio for images $B-D$). However, this factor cancels out when 
either the disk or torus components dominate the flux, so our results are 
weakly sensitive to our choice of source size (equation \ref{sourcesize}) 
and Einstein radius ($M = 0.1 M_\odot$).

\section{Conclusions}

It has long been hypothesized \citep{Sanders1989} that the near-universal 
dip near 1$\mu$m in quasar spectral energy distributions is due to the 
cutoff in emission of the dusty torus at short wavelengths due to dust 
sublimation close to the quasar.  Even if both the disk and dust emit as 
blackbodies,
the disk emits at a higher temperature and has a smaller flux than the dust 
and thus is much more compact in size, by a factor of about $\sim 100$ at 
1~$\mu$m in the rest frame.  The compact infrared disk emission should be 
more strongly affected by microlensing than the extended dusty torus 
infrared emission.  This trend is confirmed by the wavelength
dependence of the flux ratios in the IRAC bands for \qso\ (Figure 
\ref{fig:q2237fluxratios}), and is in good 
quantitative agreement with our prediction of the wavelength dependence of 
the flux ratios assuming a two-component model, accretion disk plus dusty 
torus.  Since the infrared SED of \qso\ looks very similar to a standard quasar
and similar to some low redshift Seyferts (Figures \ref{fig:q2237_netzer} 
and \ref{fig:seyferts}), this result confirms the model that quasars 
contain an accretion disk and dusty torus.   Indeed, a similar behavior
of the flux ratios was found for the two-image lensed quasar HE 1104-1805
by \citet{Poindexter2007} with optical, near-infrared and {\it Spitzer} 
observations:
in the infrared the microlensing anomalies disappear.  \citet{Poindexter2008} 
modeled the source size for this quasar as a power-law with wavelength
rather than with a two-component model;  due to the lack of variability
at the {\it IRAC} wavelengths their derived size of $\sim 3\times 10^{17}$ cm
is actually a lower limit on the source size, and thus is consistent with
a dusty torus model.  Unified models
for active galaxies \citep[e.g.][]{Antonucci1993} require a dusty torus for
obscuration of low-polarization Type I AGN to create higher polarization 
Type II AGN, while our result provides additional evidence for the 
unified model in an unobscured quasar.

Recently \citet{Kishimoto2008} have demonstrated 
the co-existence of the accretion disk and dusty torus components
near one micron using infrared spectropolarimetry.  In total flux the 
accretion disk component is masked by the stronger unpolarized thermal 
dust emission at wavelengths longer than $\sim 1\mu$m.  Since the accretion 
disk is emitted from a small scale, it can be polarized by
scattering off of gas within the dusty torus,  while the thermal emission
from the dusty torus is unpolarized since it is exterior to the scattering
region.  \citet{Kishimoto2008} have detected a polarized component with a  
power-law shape which extends into the infrared, which they identify
with polarized emission from the accretion disk, thus demonstrating
the contribution from both the disk and torus near the one micron dip.
Our results provide a complementary confirmation of the
results of \citet{Kishimoto2008}.

There are two primary areas which require improvement over the current
work: (1) time-dependent monitoring at a broad range of wavelengths
to derive the relative size versus wavelength from the microlensing, 
rather than deriving a size versus wavelength from the SED and then
predicting the microlensing behavior as we have done; 
(2) computing physically complete dusty torus SEDs coupled to accretion
disk models.

\section{Acknowledgements}

This work is based in part on observations made with the \emph{Spitzer 
Space Telescope}, which is operated by the Jet Propulsion Laboratory,
California Institute of Technology under a contract with NASA.
Support for this work was provided by NASA through an award issued by 
JPL/Caltech.  EA acknowledges the hospitality of the Institute for Theory 
and Computation at the Harvard-Smithsonian Center for Astrophysics where
a portion of this work was completed.
The authors would like to thank Ski Antonucci, \u{Z}eljko Ivezi\'{c}, Chris
Kochanek, and Julian Krolik, for helpful
discussions. We thank Chris Kochanek for providing a deconvolved $H$-band
image from the CASTLES database and Jacopo Fritz for providing
the grid of dusty torus models from his paper.  We thank the OGLE 
collaboration for making their data on Q2237+0305 available on their 
web site.  We thank Joachim Wambsganss for sharing his code for
simulating microlensing by random star fields and a helpful referee
report.  We thank Alex
Eigenbrod for sharing the wavelength dependent flux ratios measured
with VLT.  We thank Cathy Trott for sharing her model prediction for
the flux ratios prior to publication.  We thank Bruce Draine for making
his ISM dust opacity model available on his web site.
\bibliography{einsteincross02}

\end{document}